\documentclass{PoS}
\usepackage{graphics}
\usepackage{epsfig}
\title{Small $x$ physics and hard QCD processes at LHC}

\ShortTitle{Small $x$ physics and hard QCD processes at LHC}

\author{Nikolay P. Zotov\\
        SINP, Lomonosov Moscow State University\\
        E-mail: \email{zotov@theory.sinp.msu.ru}}

\abstract{We investigate the inclusive and jet production rates of $b-$quarks and also the $J/\psi-$meson production at the LHC in the framework of $k_T$-factorization QCD approach. Our study is based on the off-shell partonic QCD subprocesses. The unintegrated quark densities in a proton are determined using the Kimber-Martin-Ryskin (KMR) prescription as well as CCFM evolution equation. Our predictions are compared with the recent experimental data taken by the ATLAS, CMS  and LHCb collaborations.}

\FullConference{XX International Workshop on High Energy Physics and Quantumn Field Theory, QFTHEP 2011\\
September 24 - October 1, 2011\\
Sochi, Russia}

\begin{document}
\section{Itroduction}
 The so-called small $x$ regime of QCD is
the kinematic region, where the characteristic hard scale
of the process 
$\mu^2 \,\,\sim p_T^2 \,\,\sim M_T^2 = M^2 + p_T^2$,\,\,( $M$ and $p_T^2$ are the mass and the
transverse momentum of the final state) 
 is large as compared to the $\Lambda_{QCD}$ but $\mu$
is much less than the total c.m.s. energy $\sqrt s$ of the process:
$\Lambda_{QCD}\,\, \ll \,\, \mu \,\,\ll\,\,\sqrt s$.

In this sense, the 
HERA was the first small $x$ machine, and
the LHC is more of a small $x$ collider.
Typical $x$ values probed at the LHC in the
central rapidity region are almost two orders of
magnitude smaller than $x$ values probed at the HERA
at the same scale. Hence, the small $x$ corrections
start being relevant even for a final state with
a characteristic electroweak scale $M \sim 100$ GeV.

 It means the pQCD expansion any observable
quantity in $\alpha_s$ contains large coefficients\\
$(\ln^n(S/M^2)) \sim(\ln^n(1/x))$ (besides the usual renorm group ones $(\ln^n(\mu^2/\Lambda^2_{QCD})) $.
 The resummation of these terms
$(\alpha_s(\ln(1/x))^n \,\, \sim 1$ at $x \to 0$) in the framework of the Balitsky-Fadin-Kuraev-Lipatov (BFKL) theory~\cite{BFKL} results in the so called unintegrated gluon distribution ${\cal F}(x,{\mathbf k}_T^2)$. The unintegrated gluon distribution determines the probability to find a gluon carrying the longitudinal momentum fraction $x$ and transverse momentum $k_T$.  
This generalized factorization is called 
"$k_T-$factorization"~\cite{GLR, LRSS}. If the terms
proportional to $\alpha_s^n ln^n (\mu^2 /\Lambda^2_{QCD})$ and $\alpha_s ln^n (\mu^2 /\Lambda^2_{QCD} ) ln^n (1/x)$ are also resummed, then the
unintegrated gluon distribution function depends also on the probing scale $\mu$. This quantity depends on more degrees of freedom than the usual collinear parton density, and is therefore less constrained by the experimental data.
Various approaches to model the unintegrated gluon distribution have been proposed. One
such approach, valid for both small and large $x$, has been developed by Ciafaloni, Catani, Fiorani and Marchesini, and is known as the CCFM model~\cite{CCFM}. It introduces
angular ordering of emissions to correctly treat gluon coherence effects. In the limit of asymptotic energies, it is almost equivalent to BFKL~\cite{BFKL}, but also similar to the collinear (DGLAP) evolution for large $x$ and high $\mu^2$. The resulting unintegrated gluon distribution functions
depend on two scales, the additional scale $\bar q$ being a variable related to the maximum angle allowed in the emission.

 The BFKL evolution equation
predicts rapid growth of gluon density ($\sim x^{-\Delta}$, where $1 + \Delta$
is the intercept of so-called hard BFKL Pomeron).
However it is clear that this growth cannot continue for ever, because it would violate the unitarity constraint~\cite
{GLR}.
Consequently, the parton evolution dynamics must change at some point, and new phenomenon must come into play.
Indeed as the gluon density increases, non-linear parton
interactions are expected to become more and more important,
resulting eventually in the slowdown of the parton density growth (known as "saturation effect")~\cite{GLR, MQ}.
The underlying physics can be described by the non-linear
Balitsky-Kovchegov  (BK) equation~\cite{BK}.These nonlinear interactions lead to an equilibrium-like system of partons with some definite value
of the average transverse momentum $k_T$ and the corresponding saturation scale $Q_s(x)$.
This equilibrium-like system is the so called Color Glass
Condesate (CGC)~\cite{CGC}.
 Since the saturation scale increases with decreasing of $x$:
$Q^2_s(x,A) \sim x^{-\lambda}A^{\delta}$ (A is an atomic number )
with $\lambda \sim 0.3, \delta \sim 1/3$~\cite{Ian},
one may expect that the saturation effect will be more clear 
 at LHC energies.
\section{Unintegrated parton distributions (uPDF or TMD)}
The basic dynamical quantity in the small $x$ physics
 is  transverse-momentum-dependent (TMD) (${\mathbf k}_T$-dependent) or  unintegrated parton distribution (uPDF)
${\cal A}(x,{\mathbf k}_T^2,\mu^2)$.
For example to calculate the cross sections of photoproduction process the uPDF
${\cal A}(x,{\mathbf k}_T^2,\mu^2)$ has to be 
convoluted with the relevant partonic cross section $\hat \sigma_{\gamma g}$:
\begin{eqnarray}
  \sigma =  \int {dz\over z} \int d{\mathbf k}_{T}^2 \,
 \hat\sigma_{\gamma g}(x/z,{\mathbf k}_{T}^2,\mu^2) {\cal A}(x,{\mathbf 
k}_{T}^2,\mu^2). 
\end{eqnarray}
 For the uPDF there is no unique definition, and as a cosequence it is for the phenomenology of these quantities very important to identify uPDF which are used 
in description of high energy processes. For a general introduction to small $x$ physics and the
 small $x$ evolution equations, as well as tools for calculation in terms of MC programs, we refer to the reviews
~\cite{Small-x}. 
 
During roughly the last decade, there has been steady progress
toward a better understanding of the $k_T$-factorization (high energy factorization) and the uPDF (for example~\cite
{uPDF}.  Workshop on Transvere Momentum Distributions (TMD 2010), which was held in Trento (Italy), was dedicated to
the recent developments in small $x$ physics, based on the $k_T$-factorization and  the uPDF~\cite{TMD}.

Recently the definition for the TMDs determined
by the requirement of factorization, maximal universality and
internal consistency have been done by Collins~\cite{JC}.
   The results obtained in previous works are reduced to
the following: $k_T$(TMD)-factorization is valid in
\begin{itemize}
\item Back-to-back hadron or jet production in $e^+e^-$-annihilation,
\item Drell-Yan process ($ P_A + P_B \to (\gamma^*, W/Z) + X$),
\item Semi-inclusive DIS ($e + P \to e + h + X$).
\end{itemize}
In hadroproduction of back-to-back jets or hadrons (
$h_1 + h_2 \to H_1 + H_2 + X$)  TMD-factorization is problematic.\\
For expample, partonic picture gives the following $q_T$-dependent hadronic tensor for DY cross section:
\begin{eqnarray}
W^{\mu \nu} = \Sigma_f |H_f (Q; \mu_R)|^{\mu \nu}\\ \nonumber
  \int {d^2{\mathbf k}_{1T} d^2{\mathbf k}_{2T} {\cal A}_{f/P_1}(x_1, k_{1T}; \mu_R; {\zeta}_1) {\overline{\cal A}}_{{\overline f}/P_2}(x_2,k_{2T};\mu_R;{\zeta}_2)\delta ({\mathbf k}_{1T} + {\mathbf k}_{2T} - {\mathbf q}_T)} + Y(Q, q_T). 
\end{eqnarray}

The hard part $H_f (Q; \mu_R)$ is calculable to arbitrary order in $\alpha_s$, $\mu_R$ - the renormalization scale. The term 
$Y (Q, q_T)$ describes the matching to large $q_T$, where the approximations of TMD-factorization break down. 
The scales $\zeta_1, \zeta_2$ are related to the regulation of light-cone divergences and $\zeta_1 \zeta_2=Q^4$. The soft factors connected with soft gluons
are contained  in the definitions of the TMDs, which 
cannot be predicted from the theory and must be fitted to data.
\section{The $k_T$-factorization approach in hadroproduction}
The $k_T$-factorization approach in hadroproduction is based on the work by Catani, Ciafaloni and Hautman (CCH) (see~\cite{LRSS})
The factorization formula for $pp$-collision in physical gauge ($nA = 0, n^{\mu}= aP_1^{\mu} +
 bP_2^{\mu}$) is
\begin{eqnarray}
\sigma = {1\over 4M^2} \int d^2{\mathbf k}_{1T} \int {dx_1\over x_1} \int d^2{\mathbf k}_{2T} \int {dx_2\over x_2} {\cal F}(x_1,{\mathbf k}_{1T})
 \hat\sigma_{gg}(\rho /(x_1 x_2),{\mathbf k} _{1T},{\mathbf k}_{2T}) {\cal F}(x_2,{\mathbf k}_{2T}),
\end{eqnarray}
where $\rho = 4M^2/s$, $M$ is the invariant mass of heavy quark, and $\cal F$ are the unintegrated gluon distributions, definded by the BFKL equation:
\begin{eqnarray}
{\cal F}(x,{\mathbf k}; Q_0^2) = {1\over\pi}\delta (1 - x)\delta({\mathbf k}^2 - Q_0^2) +\\ \nonumber 
+ \overline\alpha_s \int {d^2{\mathbf q}\over{\pi\mathbf q}^2} \int {dz\over z} 
[{\cal F}(x/z, {\mathbf k} +  {\mathbf q}; Q_0^2) -
{\Theta (k - q )}{\cal F}(x/z, {\mathbf k}; Q_0^2)],
\end{eqnarray}
were $\overline\alpha_s = \alpha_s N_c/\pi$. It means
that the rapidity divergencies are cut off since there are 
an implicit cuts in the BFKL formalism. 
Effectively one introduces a cuts $\zeta_1, \zeta_2$, and
then sets $\zeta_1 = x_1, \zeta_2 = x_2$ in (3.1).\\
The declaration in CCH~\cite{LRSS} that $\cal F$ is defined via the BFKL equation (3.2) means 
that the BFKL unintegrated gluon distribution reduces to the dipole gluon distribution~\cite{BGN}. The connections between different uPDF recently were analysed in~\cite{EA}.

  The procedure for resumming inclusive hard cross-sections
at the leading non-trivial order through $k_T$-factorization was used for an increasing number of processes: photoproduction ones, DIS ones, DY and vector boson production, direct photon production, gluonic Higgs production  both in the point-like limit, and for finite
top mass $m_t$. Please look, for example~\cite{Marzani}.

 The hadroproduction of heavy quarks was considered in~\cite{BE} and recently in~\cite{Ball}.
 In last paper it was shown that when the coupling runs the dramatic enhancements seen at fixed coupling, due to infrared singularities in the partonic cross sections, are substantially reduced, to the extent that they are largely accounted for by the usual NLO and NNLO perturbative
corrections. It was found that resummation modifies the 
$B$ production cross section. at the LHC by at most 
15$\%$, but that the enhancement of gluonic $W$ production may be as large 50$\%$ at large rapidities.\\
  
In our previous papers we have used the $k_T$-factorization approach to describe experimental data on:
\begin{itemize}
\item heavy quark photo- and electroproduction at HERA
\item $J/\psi$ production in photo- and electroproduction
at HERA with taking into account the color singlet and color octet states.
\item $D^*$, $D^* + jet$, $D^* + 2jet$ photoproduction 
 and $D^*$ production in DIS
\item charm contribution to the structure function $F_2^c(x, Q^2), F_L^c, F_L$
\item $B$-meson and $b\bar b$ pair production at the Tevatron
\item charm, beauty, $D^*$ and $J/\psi$ production 
in two-photon collisions at LEP2
\item Higgs production at the Tevatron and LHC
\item  prompt photon production at the HERA and Tevatron
\item W/Z production at the Tevatron
\end{itemize}
\newpage
Here I want to present the results of
$b$-quark and $J/\psi$ production at the LHC~\cite{JKLZ, BLZ}in comparison with first experimental data obtained by ATLAS, CMS and LHCb Collaborations. The description of prompt photon production and DY lepton pairs was  done by M. Malyshev
\cite{Mal}.
\section{Ingredients of our $k_T$-factorization numerical
calculations}  
To calculate the cross section of any physical process in the $k_T-$factorization approach according to the formula
(3.1) the partonic cross section $\hat\sigma$ has to be taken off mass  shell (${\mathbf k}_{T}$-dependent) and the polarization density matrix of initial gluons has to be taken
in the so called BFKL form \footnote {The problem of choicing of  proper gauge will be discussed in more details in Sec. 5.} :
\begin{eqnarray}
 \sum \epsilon^{\mu} \epsilon^{*\,\nu} = {k_T^{\mu} k_T^{\nu} \over{\mathbf k}_T^2}.
\end{eqnarray}

Concerning the uPDF in a proton, we used two different sets.
First of them is the KMR one. The KMR approach represent an approximate treatment  of the parton evolution
mainly based on the DGLAP equation and incorpotating  the BFKL effects at the last step of the parton ladder only, in the form of the properly defined Sudakov formfactors 
$T_q({\mathbf k}_T^2,\mu^2)$ and 
$T_g({\mathbf k}_T^2,\mu^2)$, including logarithmic loop
corrections~\cite{KMR}:
\begin{eqnarray}
\displaystyle {\cal A}_q(x,{\mathbf k}_T^2,\mu^2) = T_q({\mathbf k}_T^2,\mu^2){\alpha_s({\mathbf k}_T^2)\over 2\pi}
\times\atop
{\displaystyle \times \int\limits_x^1 dz \left[P_{qq}(z) {x\over z} q\left({x\over z},{\mathbf k}_T^2\right) \Theta\left(\Delta - z\right) + P_{qg}(z) {x\over z}
 g\left({x\over z},{\mathbf k}_T^2\right) \right]},
\label{KMR_q}
\end{eqnarray}
\begin{eqnarray}
\displaystyle {\cal A}_g(x,{\mathbf k}_T^2,\mu^2) = T_g({\mathbf k}_T^2,\mu^2)
{\alpha_s({\mathbf k}_T^2)\over 2\pi} \times \atop {
\displaystyle \times \int\limits_x^1 dz \left[\sum_q P_{gq}(z) {x\over z}
q\left({x\over z},{\mathbf k}_T^2\right) + P_{gg}(z) {x\over z} g\left({x\over z},
{\mathbf k}_T^2\right)\Theta\left(\Delta - z\right) \right]},
\label{KMR_g}
\end{eqnarray}
where $\Theta$-functions imply the angular-ordering constraint $\Delta =\mu/(\mu+k_T)$ specifically to the last evolution step (to regulate the soft gluon singularities). For other evolution steps the strong ordering in transverse momentum within DGLAP equation automatically ensures angular ordering. $T_a({\mathbf k}_T^2,\mu^2)$ - the probability of evolving from ${\mathbf k}_T^2$ to $\mu^2$ without parton emission.
$T_a({\mathbf k}_T^2,\mu^2)=1$ at ${\mathbf k}_T^2 > \mu^2$.
Such definition of the ${\cal A}_a(x,{\mathbf k}_T^2,\mu^2)$ is correct for ${\mathbf k}_T^2 > \mu_0^2$ only, where
$\mu_0 \sim 1$ GeV is the minimum scale for which DGLAP evolution of the collinear parton densities is valid.\\
We use the last version of KMRW uPDF obtained from DGLAP equations~\cite{KMRW}. In this case ($a(x, \mu^2) = xG$ or
$a(x, \mu^2) =xq$) the normalization condition
\begin{eqnarray}
 a(x,\mu^2) = \int\limits_0^{\mu^2} {\cal A}_a(x,{\mathbf k}_T^2,\mu^2)d{\mathbf k}_T^2
\end{eqnarray}
is satisfied, if
\begin{eqnarray} 
{\cal A}_a(x,{\mathbf k}_T^2,\mu^2)\vert_{{\mathbf k}_T^2 < \mu_0^2} = a(x,\mu_0^2) T_a(\mu_0^2,\mu^2),
\end{eqnarray}
where $T_a(\mu_0^2,\mu^2)$ are the quark and gluon Sudakov form factors. Then the uPDF  ${\cal A}_a(x,{\mathbf k}_T^2,\mu^2)$ is defined in all ${\mathbf k}_T^2$ region.

Another uPDF was obtained using the CCFM evolution equation.
The CCFM evolution equation has been solved numerically using a Monte-Carlo method~\cite{HJ}\\
According to the CCFM evolution equation the emission of gluons during the initial cascade
 is only allowed in an angular-ordered region of phase space.
The maximum allowed angle $\Xi$ related to the hard quark box sets the scale $\mu$: $\mu^2 = \hat s + {\mathbf Q}_{T}^2(=\mu^2_f)$.\\
 The unintegrated gluon distribution  are determined  by a convolution of the non-perturbative starting distribution
 ${\cal{A}}_0(x)$
and CCFM evolution denoted by
$\bar{\cal A}(x,{\mathbf k}_{T}^2,\mu^2)$:
\begin{eqnarray}
x{\cal A}(x,{\mathbf k}_{T}^2,\mu^2)~=~\int dz {\cal A}_0(z){x\over z} {\bar{\cal A}}({x\over z},{\mathbf k}_{T}^2,\mu^2), 
\end{eqnarray}
where
\begin{eqnarray}
x{\cal{A}}_0(x)~=~Nx^{p_0}(1-x)^{p_1}\exp(-{\mathbf k}_{T}^2/k^2_0).
\end{eqnarray}
 The parameters were determined in the fit to $F_2$ data.
\section{ Heavy quark production in $pp$-interaction}
The hard partonic subprocess $g^*g^*\rightarrow Q\bar Q$ is
described by the Feynman's diagrams  presented in Fig. 1. 
\begin{figure}
\begin{center}
         \epsfig{figure=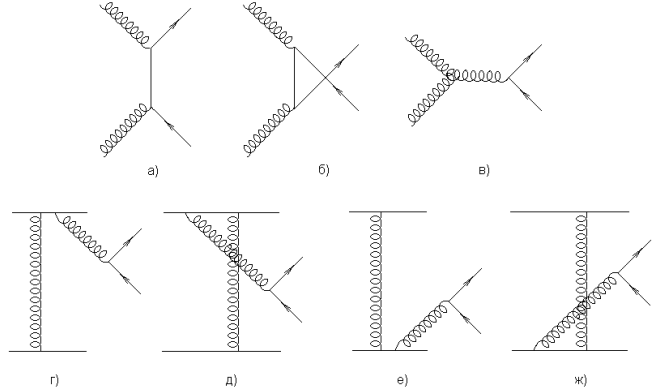,height=7cm}
\caption{\it Feynman diagrams for the $pp\rightarrow Q\bar Q
X$ process.}
\end{center}
\end{figure} 
We used Sudakov decopmosition for the momenta of heavy quarks
and the initial gluons:
$p_i = \alpha_i P_1 + \beta_i P_2 + p_{i\perp}, \,\,\, k_1 = \alpha P_1 + k_{1\perp},\,\,\,k_2 =\beta P_2 + k_{2\perp}$, 
$p_1^2=p_2^2=M^2,\,\,\,k_1^2=k_{1T}^2,\,\,\,k_2^2=k_{2T}^2$,
where in the center of mass frame of colliding particles
\begin{center}
$P_1=(E,0,0,E),\,\,\, P_2=(E,0,0,-E),\,\,\, E=\sqrt s/2,\,\,\, P_1^2=P_2^2=0,\,\,\, (P_1P_2)=s/2$.
\end{center}
Sudakov variables are
\begin{eqnarray}
\alpha_1 = \frac{M_{1T}}{\sqrt s}\exp(y_1^\ast),\,\,\,
\alpha_2 =\frac{M_{2T}}{\sqrt s}\exp(y_2^\ast),\,\,\,
\beta_1 = \frac{M_{1T}}{\sqrt s}\exp(-y_1^\ast),\,\,\,
\beta_2 = \frac{M_{2T}}{\sqrt s}\exp(-y_2^\ast),\nonumber
\end{eqnarray}
\begin{center}
$k_{1T}+k_{2T}=p_{1T}+p_{2T},\,\,\, \alpha =\alpha_1 +\alpha_2,\,\,\, \beta =\beta_1 +\beta_2$.
\end{center}

To guarantee gauge invariance, the process with off-shell incoming particles
has to be embedded into the scaterring of on-shell particles.
The second row of Fig. 1 includes non-factorizing diagrams which are factorized only in the sum. To make this factorization  one can sum up these
diagrams with the last diagram in  the first row leading to one diagram with an effective Lipatov vertex by working in Feynman gauge~\cite{Lipatov}:
\begin{eqnarray}
\Gamma^{\nu}(k_1, k_2) =  \frac{2P_1 P_2}{s}
\left(\frac{2t_1 + M^2_T}{\beta s} P_1^{\nu} - \frac{2t_2 +
 M^2_T}{\alpha s} P_2^{\nu} - (k_{1T} - k_{2T})^{\nu}\right),
\end{eqnarray}
where $t_1 =  k_1^2 = - {\mathbf k}_{1T}^2,\,\,\, t_2 =  k_2^2 = - {\mathbf k}_{2T}^2, \,\,\,
M_{T}^2 = \hat s + ({\mathbf k}_{1T}^2 + {\mathbf k}_{2T}^2)$.\\
This vertex obeys the Ward identity: $\Gamma^{\mu}(k_1, k_2)k_{1\mu}=0$.
Then the last five diagrams in Fig. 1 are replaced by one diagram (Fig. 2).
\begin{figure}
\begin{center}
         \epsfig{figure=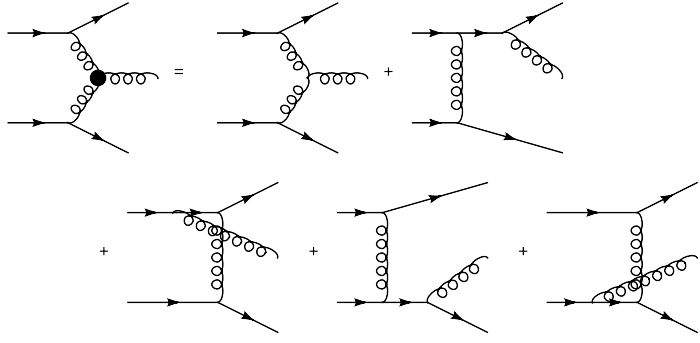,height=4.5cm}
\caption{\it Feynman diagrams explaining Lipatov vertex.}
\end{center}
\end{figure}
By neglecting the exchanged momentum in the coupling of gluons to incoming particles,
we get an eikonal vertex which does not depend on the spin of the particle:
\begin{eqnarray}
{\bar u}(\lambda_1', P_1 - k_1)\gamma^{\mu}u(\lambda_1, P_1)\,\,\,\,\,\to 2P_1^{\mu}\delta_{\lambda_1',\lambda_1}.
\end{eqnarray} 
Then it is possible to remove the external particle lines and attach so-called "non-sense" polarization to the incoming gluons: 
\begin{eqnarray}
\epsilon_{k_1}^{\mu} = {\sqrt 2}P_1^{\mu}/{\sqrt s},\,\,
\epsilon_{k_2}^{\mu} = {\sqrt 2}P_2^{\mu}/{\sqrt s}.
\end{eqnarray}
Instead of Feynman gauge, one can choose an appropriate axial gauge $n\bullet A = 0$\,\,($n^{\mu} = a P_1^{\mu} + b P_2^{\mu}$).\\
The contraction of the eikonal coupling with the gluon polarization in this gauge
\begin{eqnarray} 
d_{\mu\nu}^{(n)}(k) = - g_{\mu\nu} + \frac{n_{\mu} k_{\nu} + k_{\mu} n_{\nu}}{nk} - n^2\frac{k_{\mu}k_{\nu}}{(nk)^2}
\end{eqnarray}
then reads
\begin{eqnarray}
P_1^{\mu}d_{\mu\nu}^{(n)}(k_1) = k_{1T\nu}/\alpha,\,\,\
P_2^{\mu}d_{\mu\nu}^{(n)}(k_2) = k_{2T\nu}/\beta.
\end{eqnarray}
In such a physical gauge 
the non-factortizing diagrammes vanish since the direct 
connection of two eikonal couplings gives $P_1^{\mu}d_{\mu\nu}^{(n)}P_2^{\nu} = 0$.
It means the Lipatov vertex is to be replaced by the usual three gluon vertex. Then we can use the following matrix
elements according to the diagrams in Fig. 1:
\begin{eqnarray}
 M_1 = \bar u(p_1)(-ig\gamma^{\mu})  
\varepsilon_{\mu}(k_1)i{\hat p_1 - \hat k_1 + M\over (p_1 - k_1)^2 - M^2}(-ig\gamma^{\nu})\varepsilon_{\nu}(k_2)v(p_2),\\
M_2 = \bar u(p_1)(-ig\gamma^{\nu})
\varepsilon_{\nu}(k_2)i{\hat p_1 - \hat k_2 + M\over (p_1 - k_2)^2 -
M^2}(-ig\gamma^{\mu})\varepsilon_{\mu}(k_1)v(p_2), \\
M_3 = \bar u(p_1)C^{\mu\nu\lambda}(-k_1,-k_2,k_1+k_2)\,
{g^2\varepsilon_{\mu}(k_1)\varepsilon_{\nu}(k_2)
\over (k_1 + k_2)^2}\gamma_{\lambda}\,v(p_2),
\end{eqnarray}
where 
\begin{eqnarray}
C^{\mu\nu\lambda}(k_1,k_2,k_3)=i((k_2 - k_1)^{\lambda}
g^{\mu\nu} + (k_3 - k_2)^{\mu}g^{\nu\lambda} + (k_1 - k_3)^{\nu}g^{\lambda\mu})
\end{eqnarray}
is the standard three gluon vertex. 
\section{Numerical results}
Recently we have demonstrated reasonable  agreement between the $k_T$-factorization predictions and the Tevatron data on the $b$-quarks, $b \bar b$ di-jets, $B^+$- and  $D$-mesons
~\cite{JKLZ1}. Based on these results, here we give here
analysis of the CMS~\cite{CMS4, CMS5, CMS6} and LHCb~\cite{LHCB} data in the framework of the $k_T$-factorization approach.
We produced the relevant numerical calculations in two ways:
\begin{itemize}
\item We performed analytical parton-level calculations (which are labeled as LZ).
\item The measured cross sections of heavy quark production was compared also
with the predictions of full hadron level Monte Carlo event generator CASCADE.
\end{itemize}

In our numerical calculations we have used
three different sets, namely the CCFM A0 (B0) and the KMR ones.
The difference between A0 and B0 sets is connected with the different values of soft cut and width of the intrinsic
${\mathbf k}_{T}$ distribution.
A reasonable description of the $F_2$ data
can be achieved by both these sets.
For the input, we have used the
standard MSTW'2008 (LO)~\cite{MSTW} (in LZ calculations) and the MRST 99~\cite{MRST} (in CASCADE) sets.
The unintegrated gluon distributions  depend on
the renormalization and factorization scales $\mu_R$ and
$\mu_F$. We set 
$\mu_R^2 = m_Q^2 + ({\mathbf p}_{1T}^2 + {\mathbf p}_{2T}^2)/2$,
$\mu_F^2 = \hat s + {\mathbf Q}_T^2$, where ${\mathbf Q}_T$ is the
transverse momentum of the initial off-shell gluon pair,
$m_c = 1.4 \pm 0.1$~GeV, $m_b = 4.75 \pm 0.25$~GeV.  We use the LO formula
for the coupling $\alpha_s(\mu_R^2)$ with $n_f = 4$ active quark flavors at $\Lambda_{\rm QCD} = 200$~MeV, such
that $\alpha_s(M_Z^2) = 0.1232$.\\

We begin the discussion by presenting our results for the
muons originating from the semileptonic decays of the $b$
quarks. The CMS collaboration has measured the transverse momentum and the pseudorapidity distributions of  muons from 
$b$-decays. The measurements have been performed
in the kinematic range $p_T^\mu > 6$~GeV and $|\eta^\mu| < 2.1$ at the total center-of-mass energy $\sqrt s = 7$~TeV.\\
To produce muons from $b$-quarks, we first convert
$b$-quarks into $B$-mesons
using the Peterson fragmentation function with default value
$\epsilon_b = 0.006$
and then simulate their semileptonic decay according to the standard electroweak theory taking into account the decays 
$b \to \mu $ as well as the cascade decay $b\to c\to \mu$.\\

  The results of our calcilations are shown in Figs. 3 -- 8
in the comparison with the LHC data (see~\cite{JKLZ} for more details).
We obtain a good description of the data when using 
the CCFM-evolved (namely, A0) gluon distribution in LZ calculations.
The shape and absolute normalization of measured $b$-flavored hadron cross sections at forward rapidities are reproduced well (see Fig.~6). 
The KMR and CCFM B0 predictions are somewhat below the data.
In contrast with $b$ hadron and decay muon cross sections, 
the results for inclusive $b$-jet production 
based on the CCFM and KMR gluons are very similar to each other anda reasonable description of the data is obtained by all unintegrated gluon distributions under consideration.
 
 The CASCADE predictions tend to lie slightly below the LZ ones and are rather close to the {\textsc MC@NLO} calculations (not shown). The observed difference between the LZ and CASCADE  connects with the missing parton shower effects in the LZ evaluations.
We have checked additionaly that the LZ and CASCADE predictions coincide at parton level..

Figs. 5 and 8 show
the role of fragmentation and off-shelness effects in our calculations.

\begin{figure}
\epsfig{figure=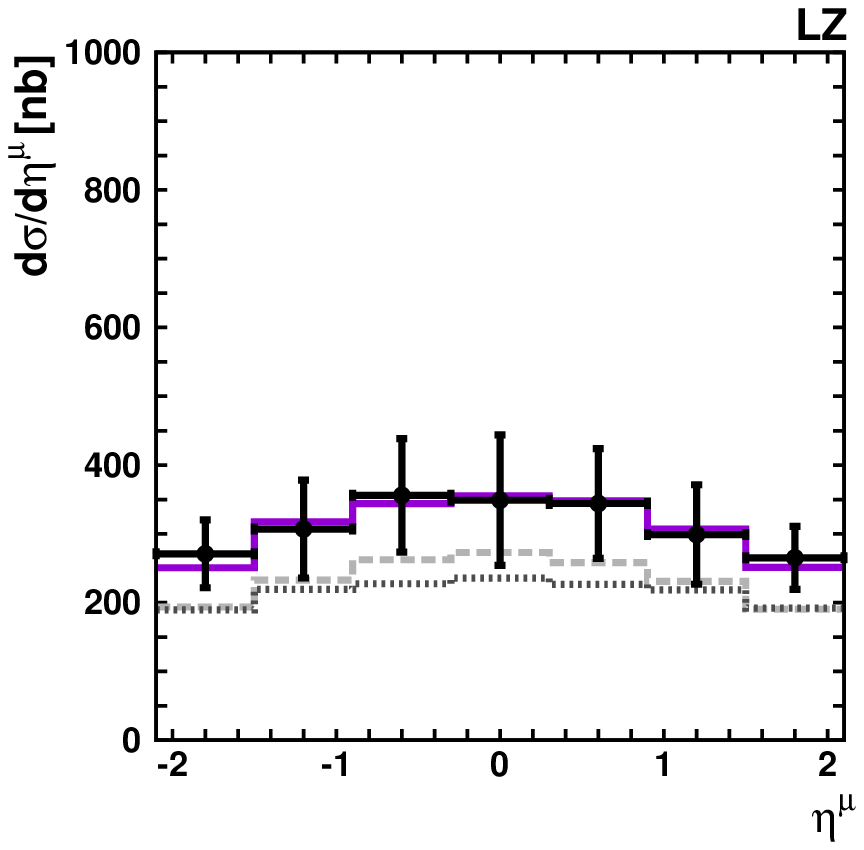, width = 8.1cm}
\put(0.0,1)
{\epsfig{figure=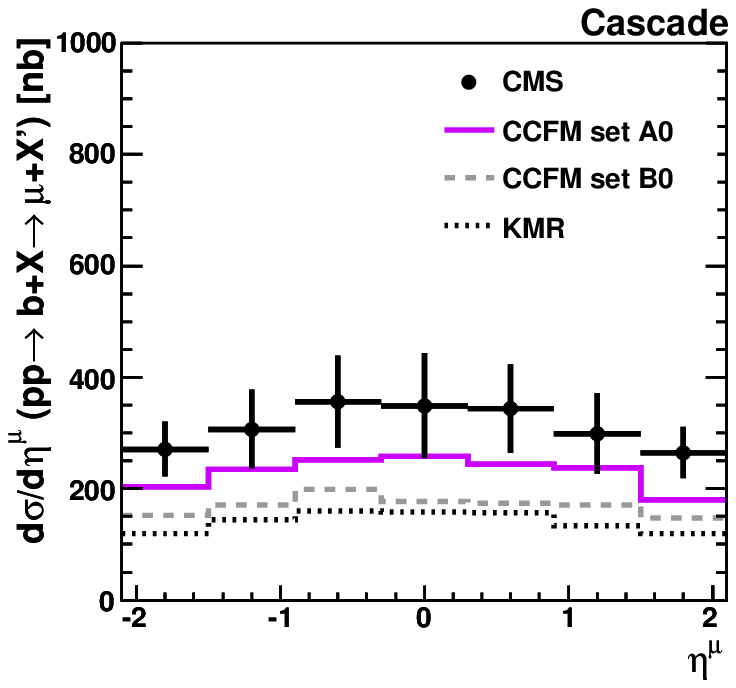, width = 7.cm}}\\
\caption {\it The pseudo-rapidity distributions of muons arising from the semileptonic decays of beauty quarks. The first column shows the LZ numerical results while the second one depicts the CASCADE predictions.
The solid, dashed and dash-dotted, dotted histograms correspond to the results obtained with the CCFM A0, B0
and KMR unintegrated gluon densities.
The experimental data are from CMS~\cite{CMS4}.}
\end{figure}
\begin{figure}
\epsfig{figure=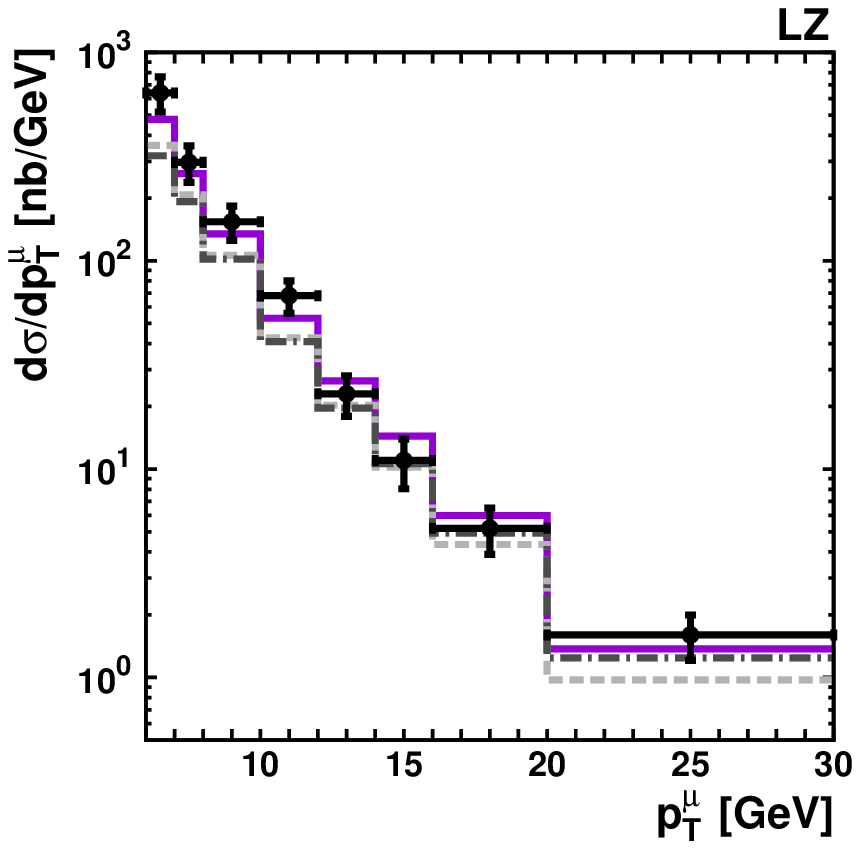, width = 8.1cm}
\put(0.0,0.5)
{\epsfig{figure=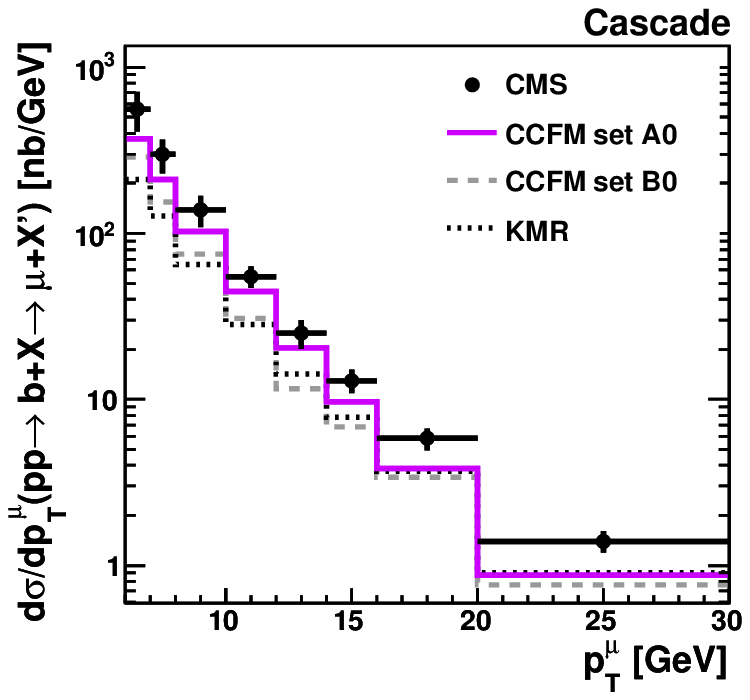, width = 7.cm}}\\
\caption{\it The transverse momentum distributions of muons arising from the
semileptonic decays of beauty quarks. The first column shows the LZ numerical
results while the second one depicts the CASCADE predictions.
Notation of all histograms is the same as on previous slide.
The experimental data are from CMS~\cite{CMS4}.}
\end{figure}
\begin{figure}
\epsfig{figure=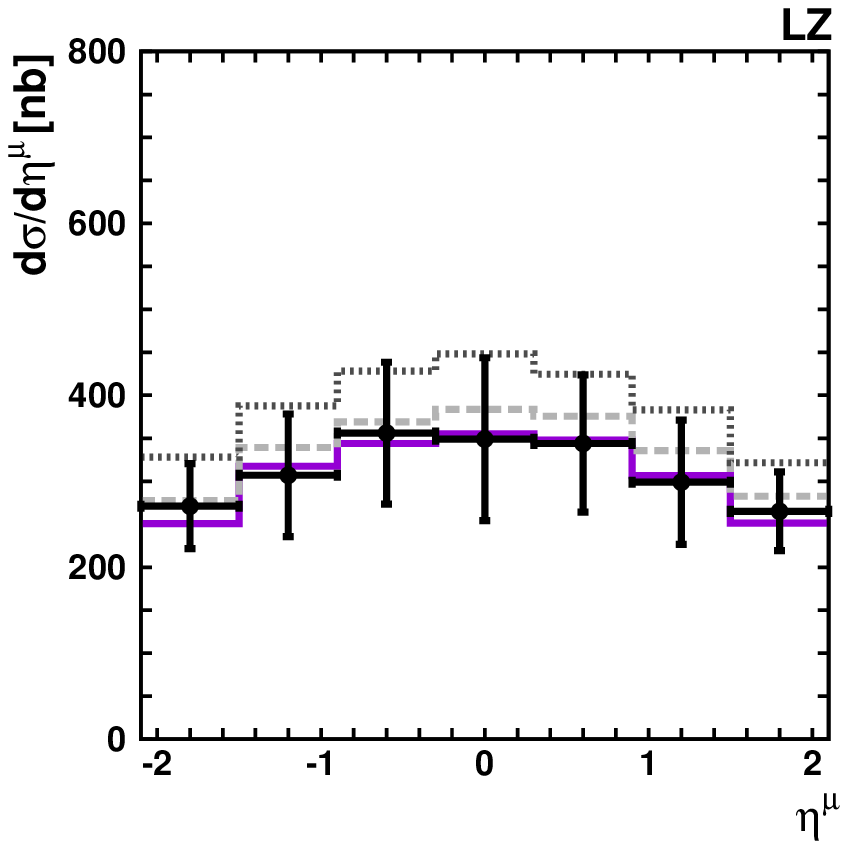, width = 8cm}
\epsfig{figure=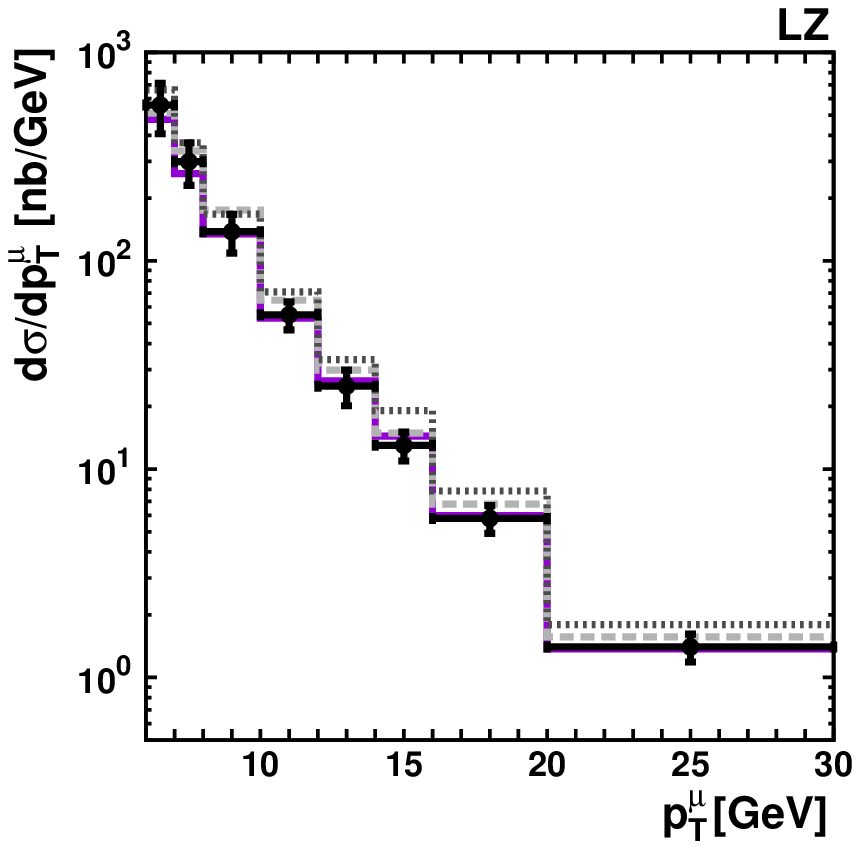, width = 8cm}\\
\caption{\it The dependence of our predictions on the fragmentation scheme.
The solid, dashed and dash-dotted histograms correspond to the results
obtained using the Peterson fragmentation function with
$\epsilon_b = 0.006, \epsilon_b = 0.003$ and the non-perturbative fragmentation functions
respectively. We use CCFM (A0) gluon density for
 illustration. The experimental data are from CMS~\cite{CMS4}.}
\end{figure}
\begin{figure}
\begin{center}
\epsfig{figure=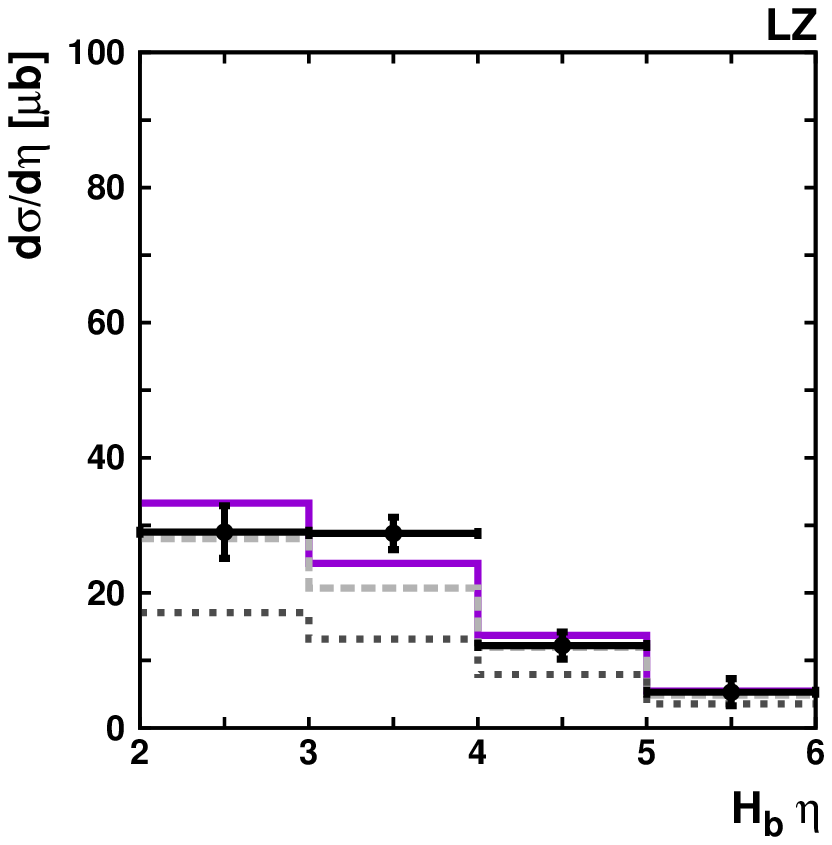, width = 8cm}
\epsfig{figure=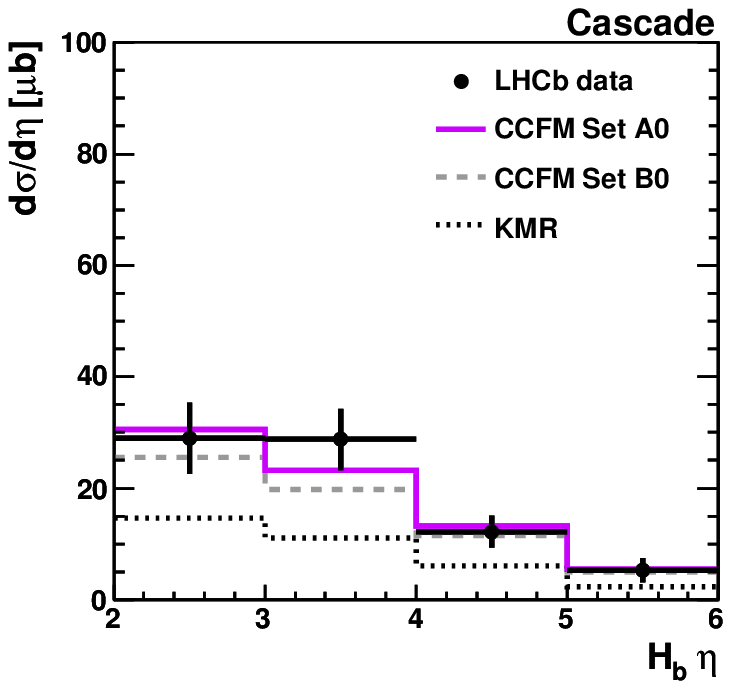,width = 7cm}
\end{center}
\caption{\it The pseudorapidity distributions of $b$-flavored
hadrons at LHC. LZ results with the CCFM AO, BO and KMR uPDF. The experimental data are from LHCb~\cite{LHCB}.}
\end{figure}
\begin{figure}
\begin{center}
\epsfig{figure=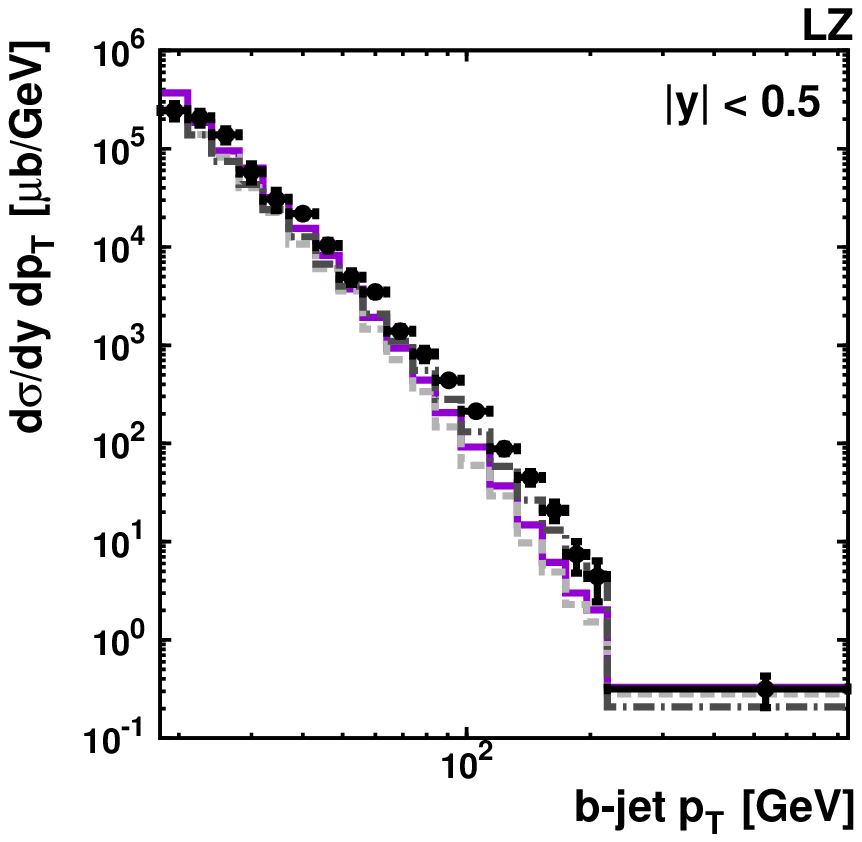, width = 6cm}
\epsfig{figure=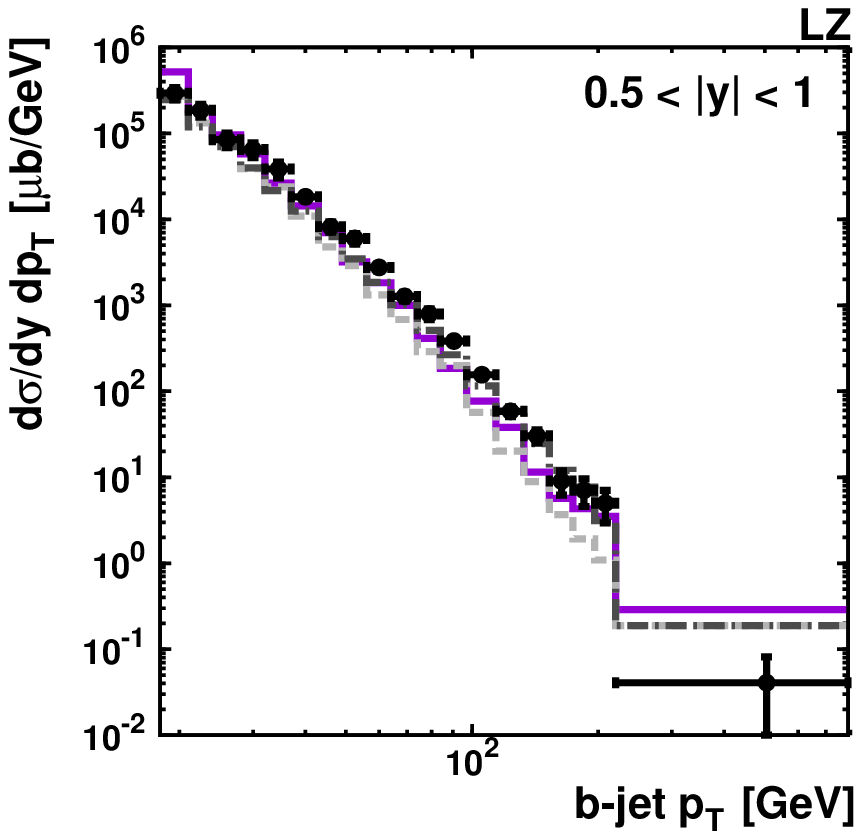, width = 6cm}
\epsfig{figure=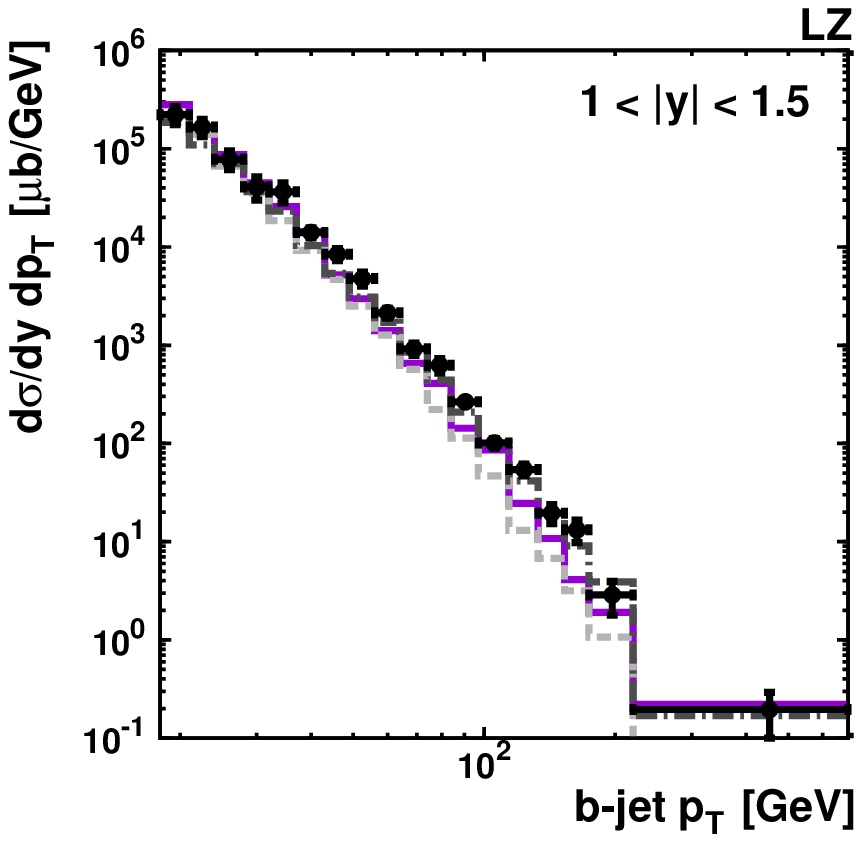, width = 6cm}
\epsfig{figure=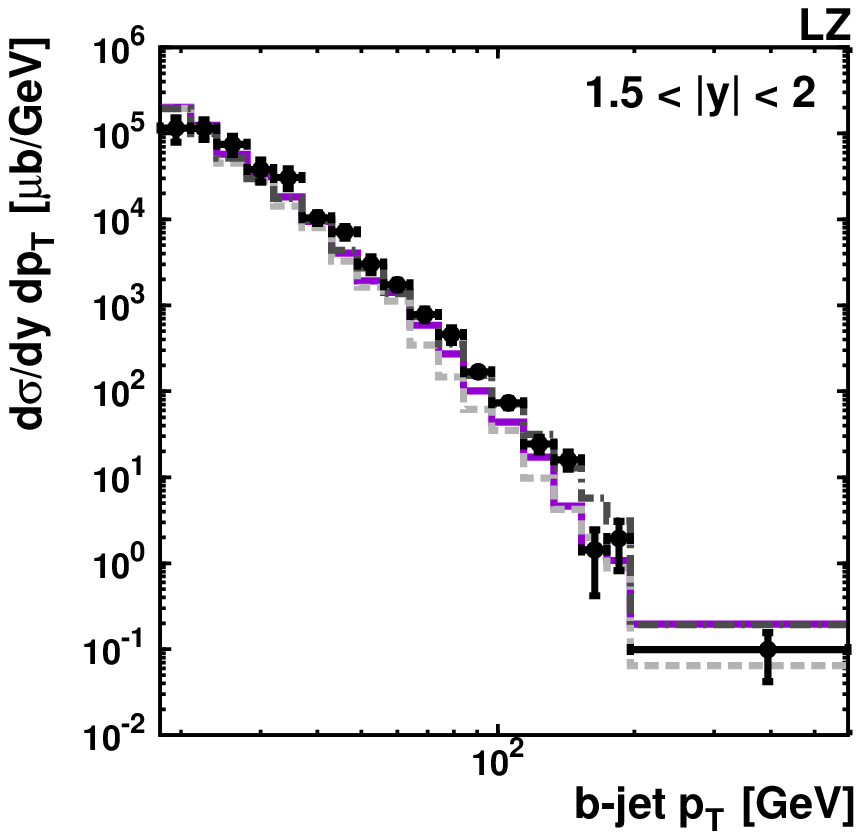, width = 6cm}\\
\caption{\it The differential cross sections 
$d\sigma/dy\,dp_T$ of inclusive $b$-jet production integrated
over the specified $y$ intervals.
The experimental data are from CMS~\cite{CMS6}}
\end{center}
\end{figure}
\begin{figure}
\begin{center}
\epsfig{figure=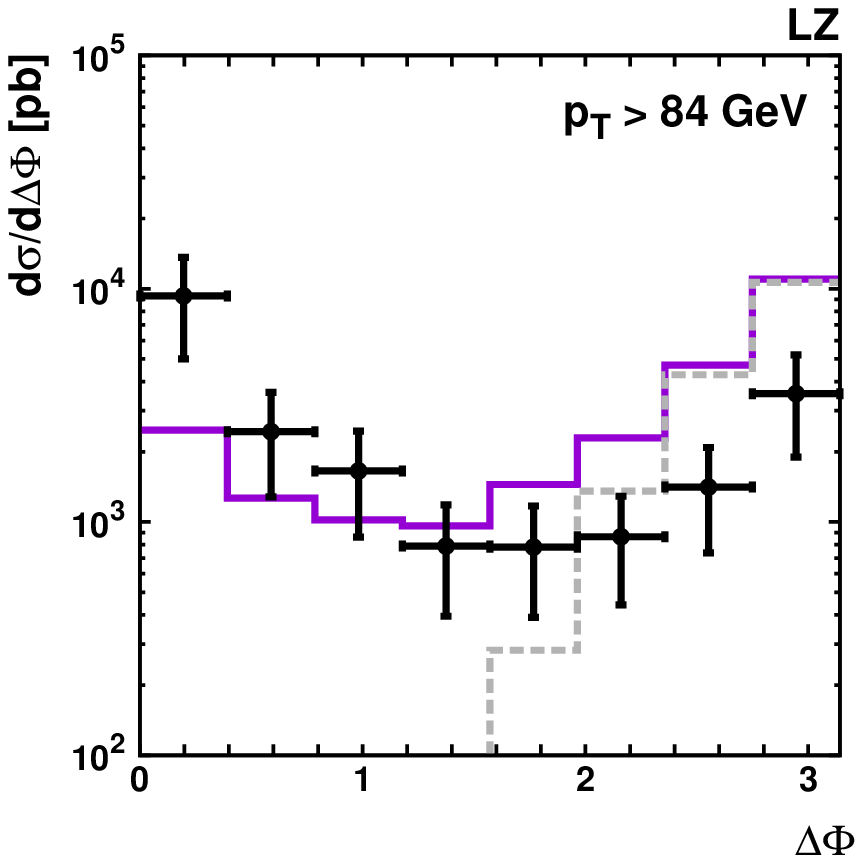, width =6.5cm}
\epsfig{figure=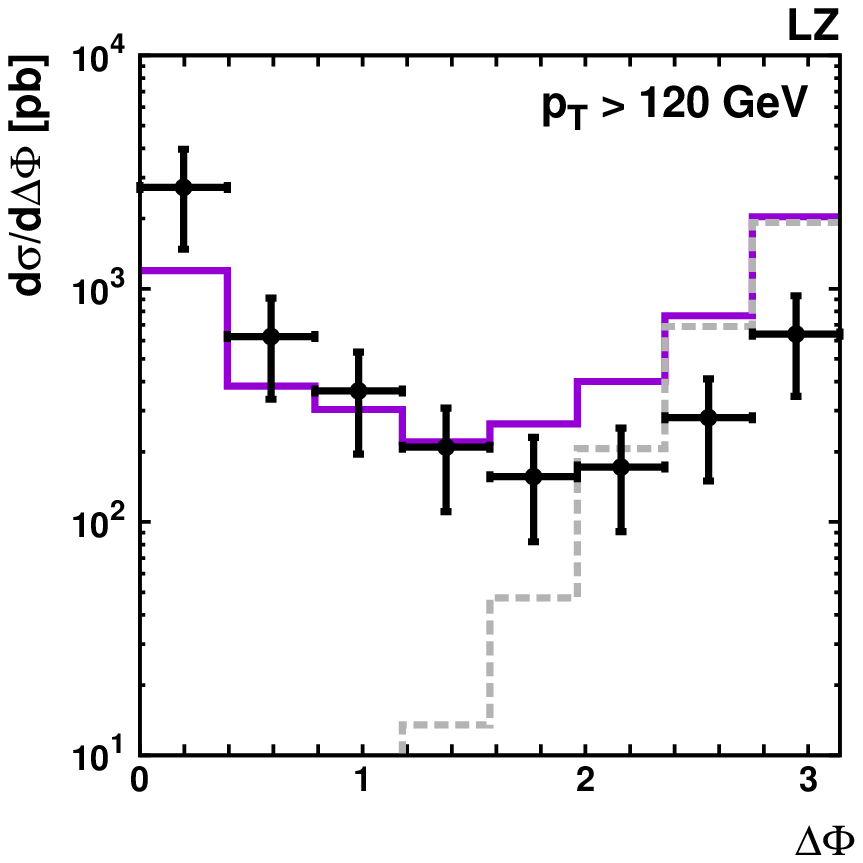, width =6.5cm}
\epsfig{figure=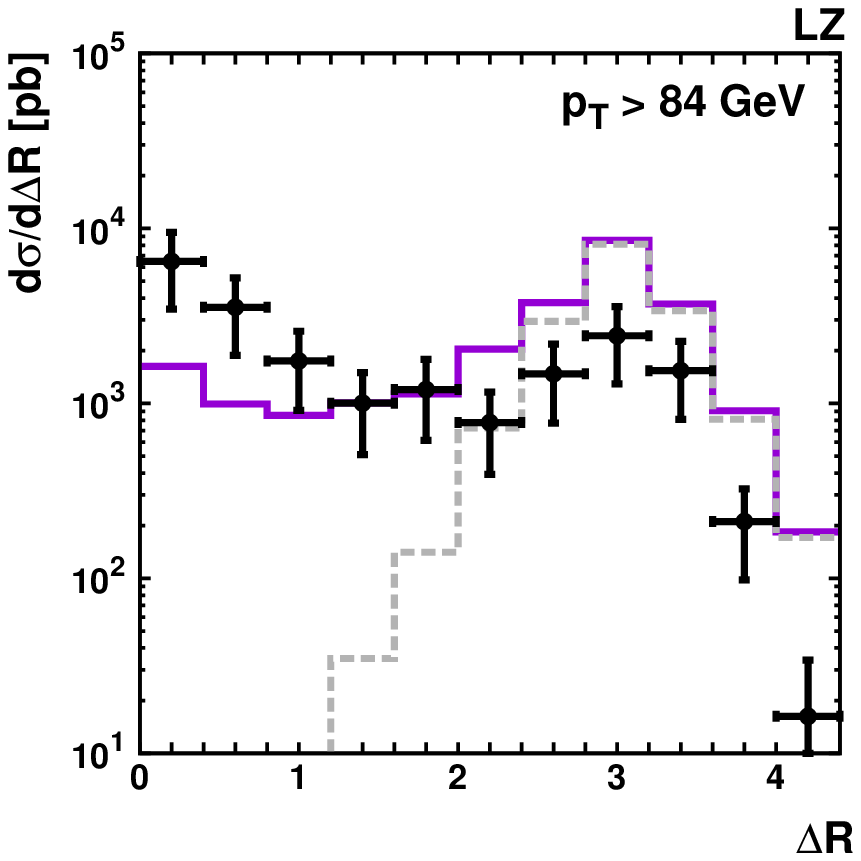, width =6.5cm}
\epsfig{figure=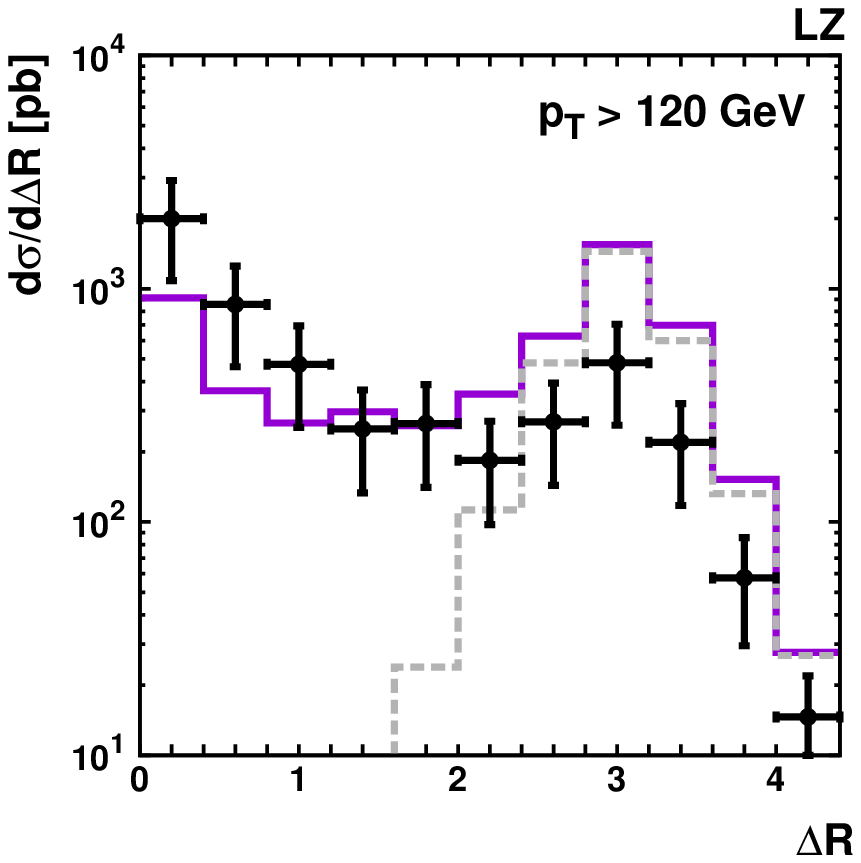, width =6.5cm}\\
\end{center}
\caption{\it Importance of non-zero ${\mathbf k}_T$ of
incoming gluons. Dotted histograms - the results obtained
without the virtualities gluons and with 
${\mathbf k}_T^2 < \mu^2_R$ in matrix element. The CMS data~\cite{CMS5}.}
\end{figure}
\section{Quarkonium production in the $k_T-$factorization approach}
The production of prompt $J/\psi$($\Upsilon$)-mesons in 
$pp$-collisions can proceed via either
direct gluon-gluon fusion or the production of $P$-wave states $\chi_c (\chi_b$) and $S$-wave state $\psi'$
followed by their radiative decays $\chi_c (\chi_b)\to J/\psi(\Upsilon) + \gamma$.
 In the $k_T-$factorization approach the direct mechanism corresponds to the partonic subprocess $g^* + g^*\to J/\psi(\Upsilon) + g$.
 The production of $P$-wave mesons is given by
$g^* + g^*\to\chi_c(\chi_b)$, and there is no emittion of any additional gluons. The feed-down contribution from $S$-wave state $\psi'$ is described by $g^* + g^*\to \psi' + g$.\\
 The cross sections charmonium states depend on the renormalization and factorization scales $\mu_R$ and
$\mu_F$. We set $\mu_R^2 = m^2 + {\mathbf p}_T^2$ and $\mu_F^2
 = \hat s + {\mathbf Q}_T^2$, where 
${\mathbf Q}_T^2$ is the transverse momentum of initial 
off-shell gluon pair. Following to PDG~\cite{PDG}, we set $m_{J/\psi} = 3.097$ GeV, $m_{\chi c1} = 3.511$ GeV, $m_{\chi c2} = 3.556$ GeV, $m_{\psi '} = 3.686$ GeV and use the LO formula for the coupling constant $\alpha_s(\mu^2)$ with 
$n_f =4$ quark flavours at $\Lambda_{QCD} = 200$
Mev, such that $\alpha(M^2_Z) = 0.1232$.\\ 
The charmonium wave functions are taken to be equal to 
$|{\cal}R_{J/\psi}(0)|^2/4{\pi} = 0.0876$ GeV$^3$,
$|{\cal}{R'}_{\chi}(0)|^2 = 0.075$ GeV$^5$,  
$|{\cal}R_{\psi '}(0)|^2/4{\pi} = 0.0391$ GeV$^3$ and
the following branching fractions are used
 $B(\chi_{c1} \to J/\psi +
\gamma) = 0.356, B(\chi_{c2} \to J/\psi + \gamma) = 0.202,
B(\psi ' \to J/\psi + X) = 0.561$ and $B(J/\psi \to
\mu^+\mu^-) = 0.0593$.
Since the branching fraction for $\chi_{c0} \to J/\psi
+ \gamma $ decay is more than an order of magnitide smaller than for $\chi_{c1}$ and $\chi_{c2}$, we neglect its contribution to $J/\psi$ production.
As $\psi' \to  J/\psi + X$ decay matrix elements are unknown, 
these events were generated according to the phase space.

Comparison the results of our calculations with the CMS~\cite{CMSJ}, ATLAS~\cite{ATLASJ} and LHCb~\cite{LHCbJ} data are shown in Figs. 9 - 11~\cite{BLZ}. We see  that the taking into
account sole direct production is not sufficient to descibe
the LHC data.We have obtained a good overall agreement between our predictions and the data when summing up the
direct and feed-down contributions. The dependence of
our numerical results on the uPDF is rather weak and the
CCFM and KMR predictions are practically coincide.  
The difference between them can be observed at small $p_T$
or at large rapidities probed at the LHCb measerements.
We have evaluated the polarizations parameters of prompt
$J/\psi$ mesons  in the kinematical region of CMS, ATLAS and
LHCb measurements in the Collins-Soper and  helicity frames
(see~\cite{BLZ}). We have took into account the contributions
from the direcct and feed-dowm mechanisms. The qualitative
predictions for the $J/\psi$ meson polarization are stable
with respect to variations in the model parameters.
Therefore future price measurements of the polarizations 
parameters of the $J/\psi$ mesons at the LHC will play
crucial role in discriminating the different theoretical
approaches.
 
\begin{figure}
\begin{center}
\epsfig{figure=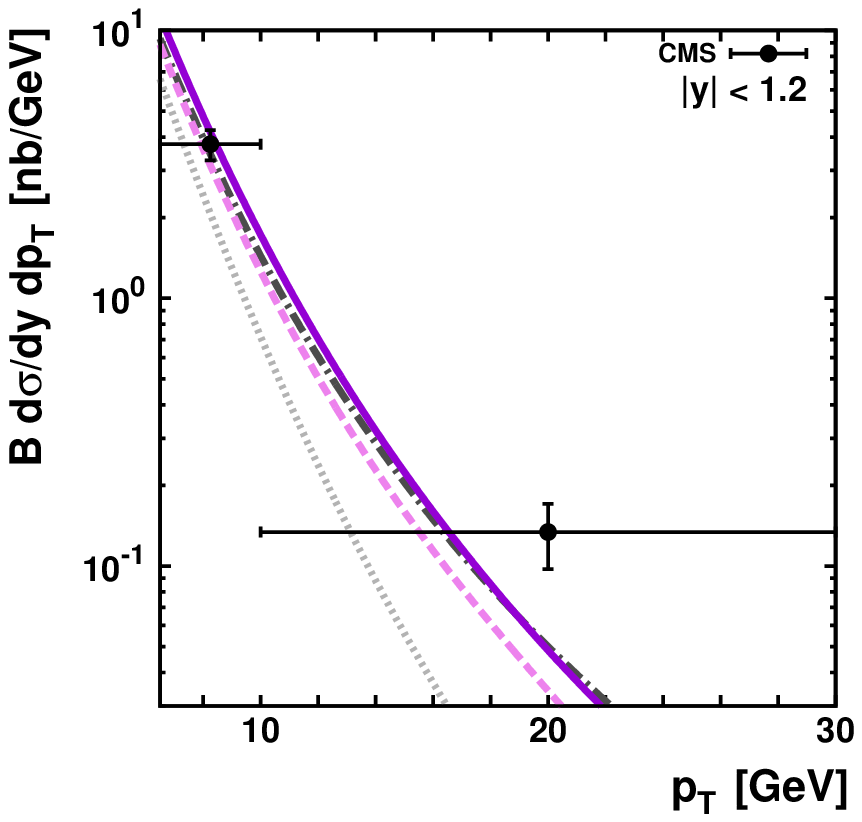, width =6.5cm}
\epsfig{figure=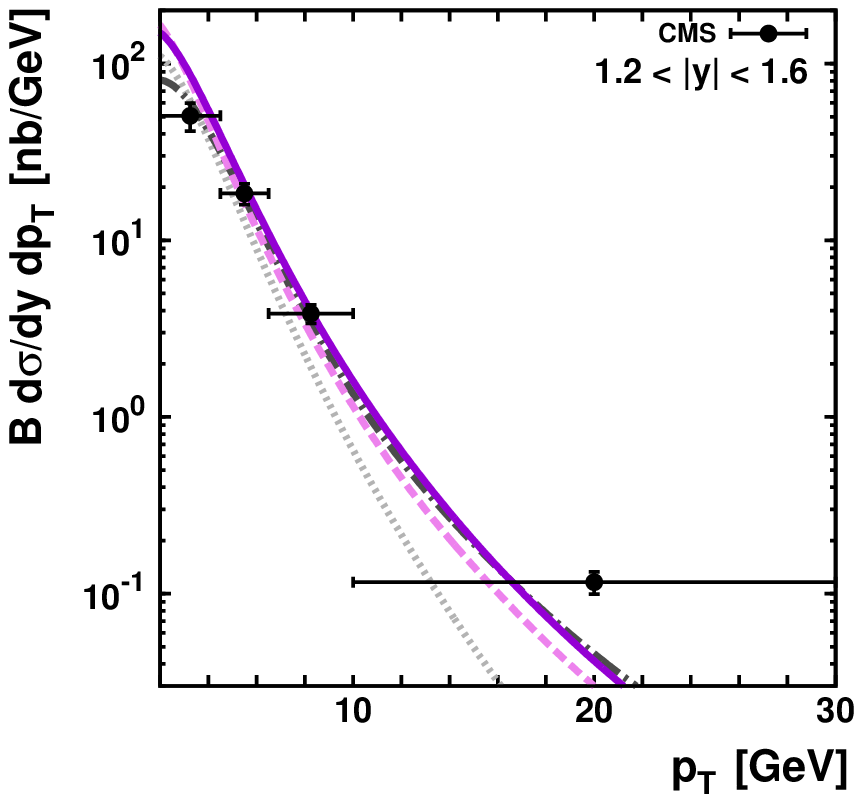, width =6.5cm}
\epsfig{figure=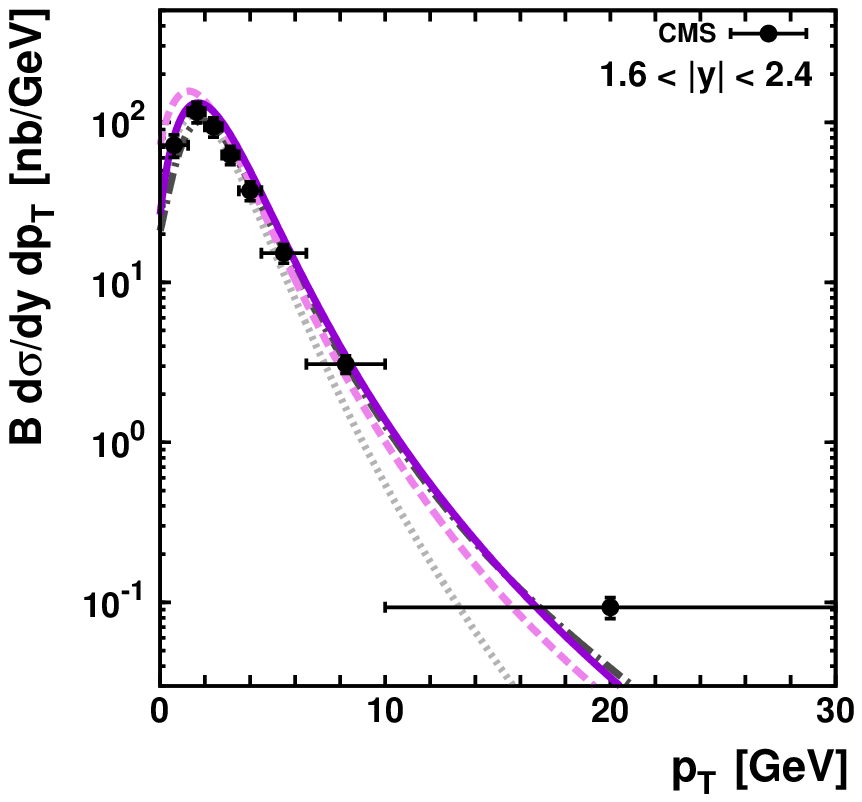, width =6.5cm}\\
\end{center}
\caption{\it The differential cross sections $d\sigma/dydp_T$ of prompt $J/\psi$ production at LHC integrated over the specified $y$ intervals. Solid, dashed and dashed-dotted curves correspond to the results obtained using the CCFM A0, B0 and KMR uPDF. Dotted curves represent the contribution from sole direct  production mechanism cal
culated with the CCFM A0 uPDF. The CMS data~\cite{CMSJ}.}
\end{figure}
%
%
\begin{figure}
\begin{center}
\epsfig{figure=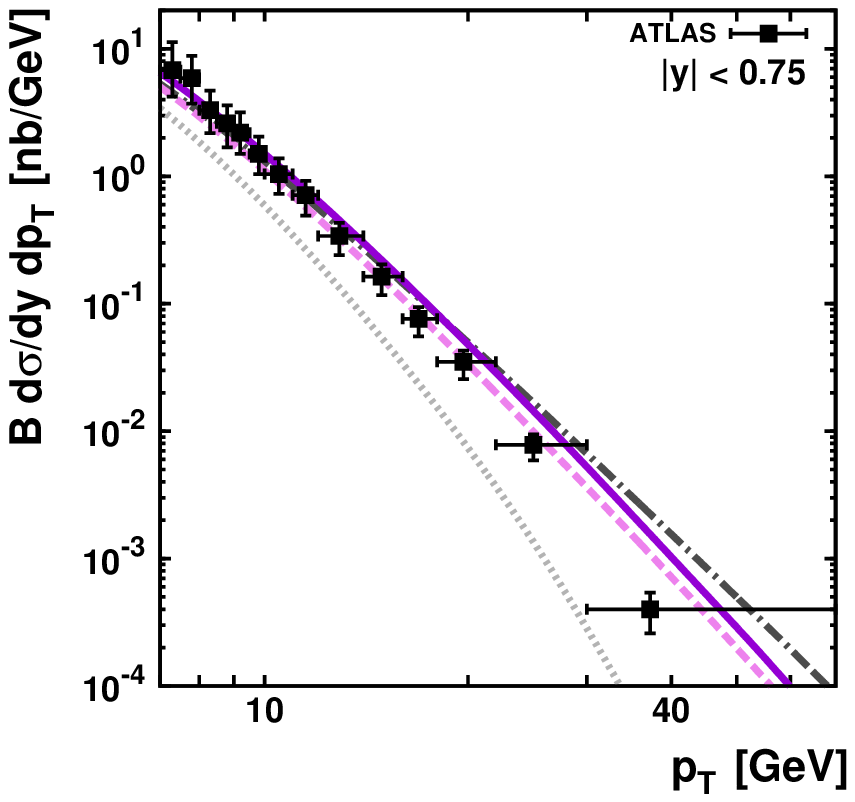, width =6cm}
\epsfig{figure=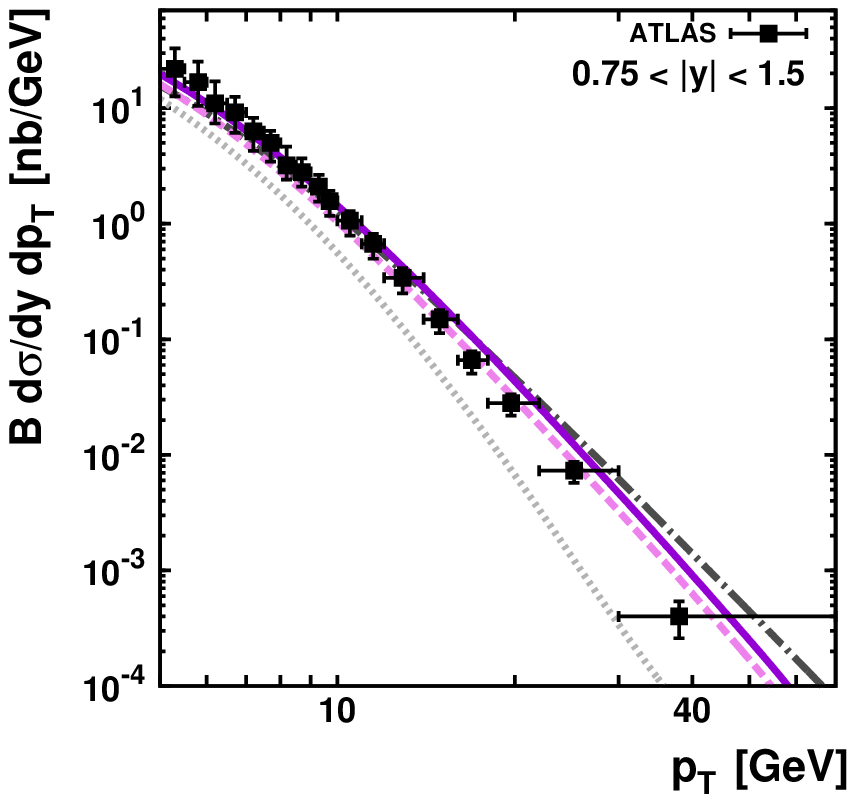, width =6cm}
\epsfig{figure=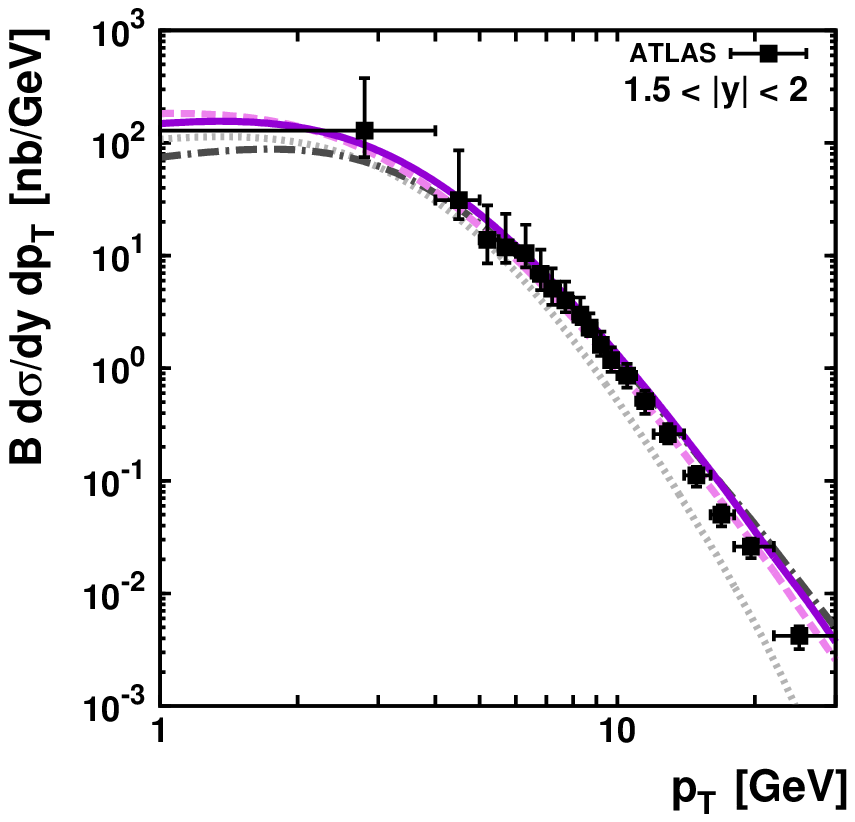, width =6cm} 
\epsfig{figure=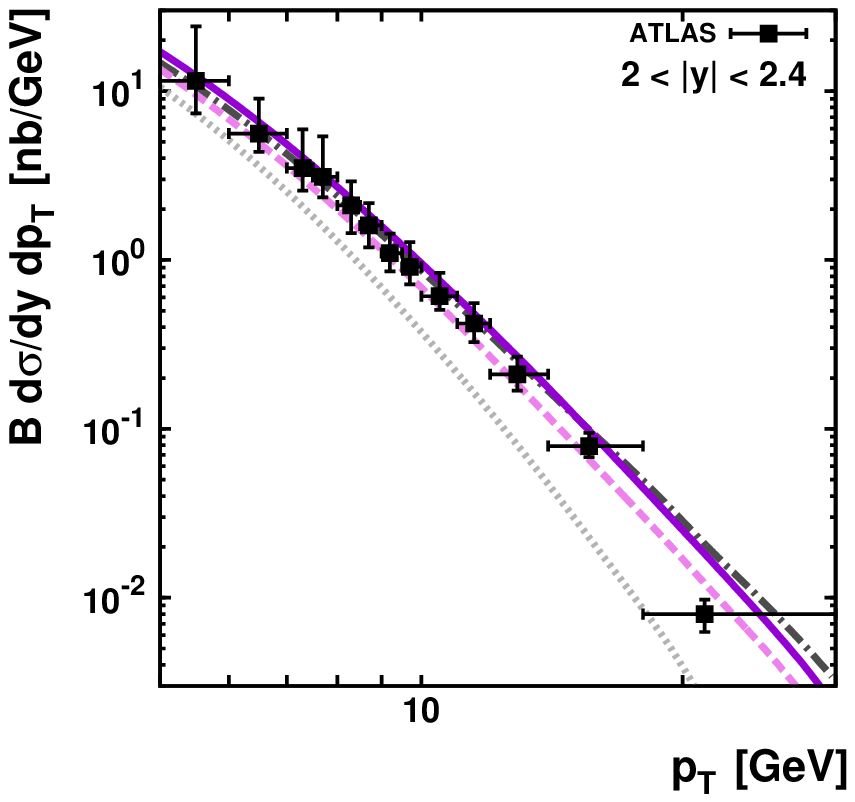, width =6cm}\\
\end{center}
\caption{\it The differential cross sections $d\sigma/dydp_T$ of the $J/\psi$ production at LHC integrated over the specified $y$ intervals in comparison with the ATLAS data~\cite{ATLASJ}.}
\end{figure}
\begin{figure}
\begin{center}
\epsfig{figure=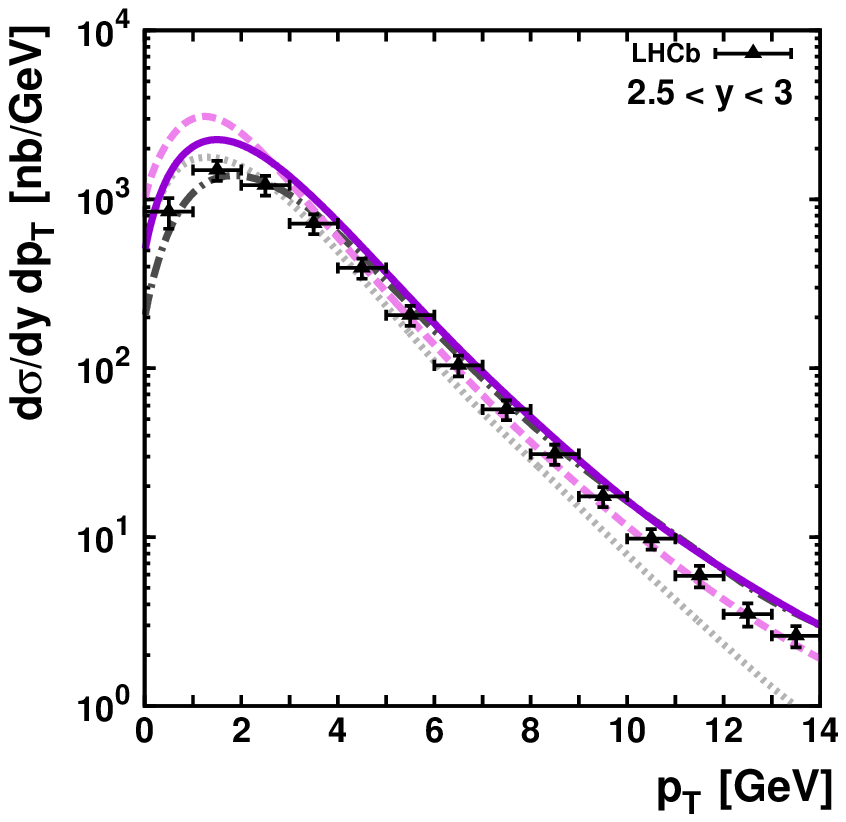, width =6cm}
\epsfig{figure=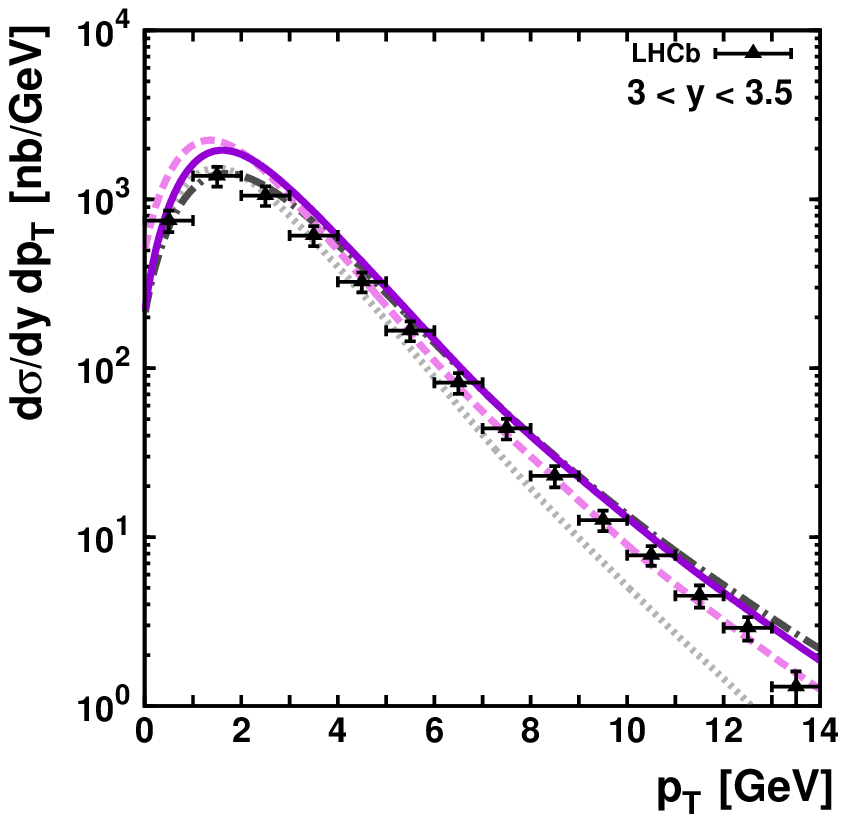, width =6cm}
\epsfig{figure=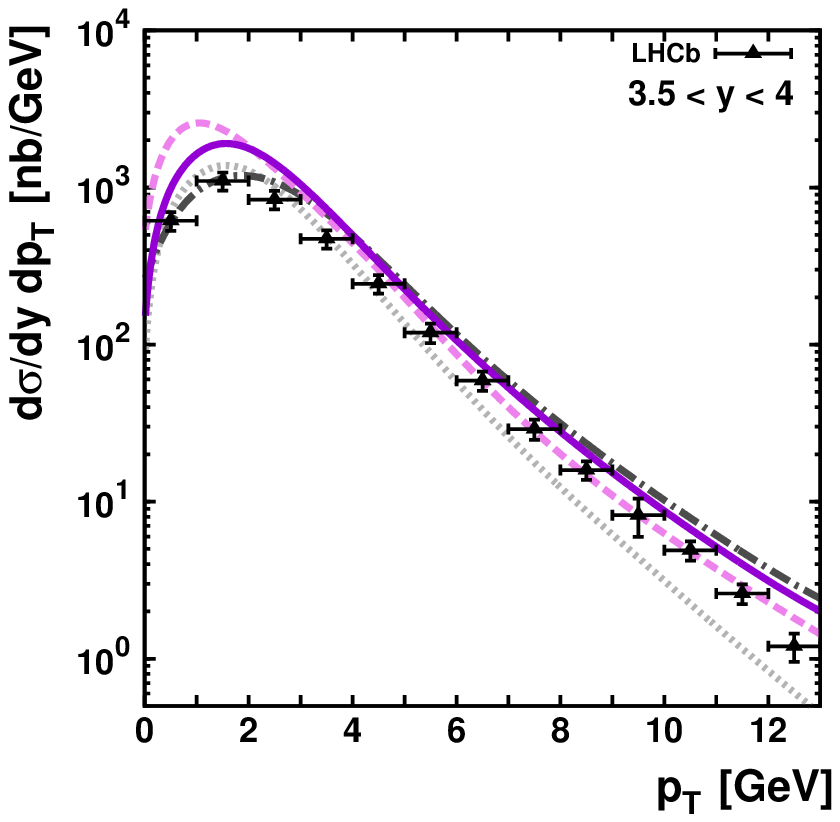, width =6cm} 
\epsfig{figure=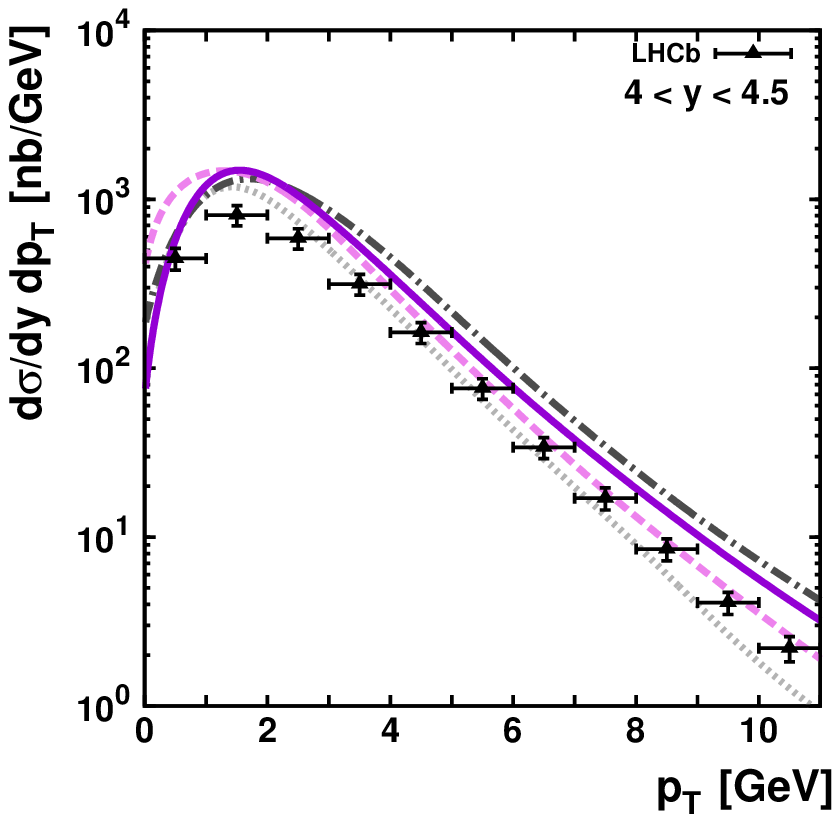, width =6cm}\\
\end{center}
\caption{\it The differential cross sections $d\sigma/dydp_T$  of the $J/\psi$ production at LHC integrated over the specified $y$ inrervals in comparison with the LHCb data~\cite{LHCbJ}.}
\end{figure}
\section{Conclusions}
  In the present time there is steady progress
toward a better understanding of the $k_T$-factorization (high energy factorization) and the uPDF (TMD).\\
 We have described the first exp. data of $b$-quark
 and $J/\psi$ production at LHC in the $k_T$-factorization approach. We have obtained reasonable agreement
of our calculations and the first experimental data taken by
the CMS and ATLAS Collaborations.\\
The dependence of our predictions on the uPDF appears 
at small transverse momenta and at large rapidities in $H_b$ and $J/\psi$ production covered by the LHCb experiment.\\
Our study has demonstrated also that
in the framework of the $k_T$-factorization approach there is no need in a color octet contributions for the charmonium production at the LHC.\\
As it was shown in~\cite{BLZ} the future experimental analyses of
quarkonium polarization at LHC are very important
and informative for discriminating the different theoretical models.
\section*{Acknowledgements}
I'd like to thank L.N. Lipatov for very useful discussion of different problems connected with subject of this talk.
I'm very grateful to S.P. Baranov, H. Jung, M. Kr\"amer and
A.V. Lipatov for fruitful collaboration. I thank S.P. Baranov for careful reading the manuscript and very useful remarks. This research was  supported by DESY Directorate in the framework of Moscow -- DESY project on Monte-Carlo implementation for HERA -- LHC, by the FASI of Russian Federation (grant NS-1456.2008.2), FASI state contract 02.740.11.0244, RFBR grant 11-02-01454-a and also by the RMES (grant the Scientific Research on High Energy Physics).

\end{document}